# Oblique rotational axis detection using elliptical optical vortex based on rotational Doppler effect

XIANGYANG ZHU[2,*], YUAN REN[1], YAOHUI FAN[2], XINYI WEN[2], XIAOCEN CHEN[3], RUOYU TANG [2], YOU DING [2], ZHENGLIANG LIU[2] AND TONG LIU[2,*]

[1]*Department of Basic Course, Space Engineering University, Beijing, 101416, China*
[2]*Department of Aerospace Science and Technology, Space Engineering University, Beijing, 101416, China*
[3]*Beijing Institute of Special Electromechanical Technology, Beijing, 100191, China*
*li_cling_yu@163.com*
*liutong719@163.com*

**Abstract:** The rotational Doppler effect (RDE) of structured light carrying orbital angular momentum (OAM) has attracted widespread attention for applications in optical sensors and OAM spectrum detection. These studies, however, based on RDE are mostly focus on the motion parameters of rotating object, other equally important attitude characteristics e.g., the tilt angle of the axis of rotation have rarely been considered. We observed an interesting phenomenon in the experiments: the rotational Doppler spectral distribution varies with the ellipticity of elliptical optical vortex (EOV) and the tilt angle between the rotational axis and optical axis, which inspired us wonder if it is possible to detect oblique rotational axis or compensate the rotational Doppler broadening effect induced by oblique incidence by utilizing the EOV. Here, we reveal the RDE quantitative relationship with tilt angle and ellipticity for the first time, and report a novel approach for tilt angle measurement. By employing a series of EOV with periodically varying ellipticity to illuminate a rotating object and analyze the time-frequency spectral distribution of scattered light associated with ellipticity and tilt angle, the tilt angle can be acquired accurately based on the specific relationship between the tilt angle and ellipticity of EOV. Furthermore, the spectrum broadening effect arising from oblique incidence in the actual scenario may be addressed through our scheme. The method may find applications in industrial manufacturing and target attitude measurement, and our results provide new insights for obtaining more information about objects.

## 1. Introduction

Rotating motion exists widely in free space. Although a great deal of research about the micro-Doppler effect has been done[1-3], it is still complex and difficult to accurately extract the parameters of the rotational motion from conventional radar echo signals. The discovery of the orbital angular momentum(OAM) of the optical vortex(OV) has brought dawn to the convenient and accurate extraction of rotational parameters of the object[4]. The vortex beam is a kind of structured light with a helical phase structure described by $\exp(i\ell\varphi)$, where $\ell$ is the TC and $\varphi$ is the azimuthal coordinate[5]. Such beams can be sensitive to the rotating motion of the objects, which will result in the frequency shift of the optical echo[6]. Different from the linear Doppler effect[7, 8], this phenomenon is known as the rotational Doppler effect(RDE), which has aroused great interest of researchers[9-11]. From rotating speed measurement[12] to OAM mode analysis[13-15] and nonlinear optics[16, 17], a large number of studies on the RDE have been conducted. In addition, it has also been applied in measurements for compound motion decoupling[18, 19] and the azimuth recognition of the rotating axis[20].

However, most of the current researches on RDE requires that the OV axis coincides with the spinning axis of the object which is difficult to satisfy in the field of engineering[19, 21,

22]. It is more general and common case that the OV illuminates obliquely the rotating object. In this condition, the rotational Doppler shift will broaden[23]. And as the tilt angle increases gradually, the spectrum broadening effect is more serious and the signal amplitude decreases ,which will lead to the rotating speed cannot be extracted finally. Moreover, in the actual telemetry, it is difficult to keep the optical axis coincident with the rotating axis due to the disturbance of atmospheric turbulence and object motion, which makes it complicated to measure the rotating speed.

To overcome this problem, a novel method for rotational Doppler shift correction based on the EOV is proposed in this paper. Firstly, inspired by the principle of geometric projection, we generate the EOV with controllable eccentricity as the probe beam via reverse design and coordinate transformation techniques. And the RDE mechanism of the EOV under the condition of oblique illumination is revealed in the study. This research uses the asymmetric optical characteristics of the EOV to compensate for the rotational Doppler spectral spreading effect caused by tilted detection condition. By this method, the signal-to-noise ratio of the echo signal can be improved, which is beneficial to accurately extract the rotating speed, increase the detection range in remote sensing, and extend the application scenario. Secondly, in the study, a set of superimposed EOVs(SEOVs) with continuously varying ellipticities is employed to illuminate the rotating object under the misaligned condition, and the time-frequency analysis method is used to process the echo signal. Through the analysis of rotational Doppler time-frequency spectrum, the tilt angle of the spinning axis of the object with respect to the optical axis and the optimal ellipticity of the EOV can be obtained simultaneously. In addition, considering the propagation characteristics of the EOV[24], we can modify the rotational angle of the probe beam in advance to improve the mode purity of the EOV in telemetry. At last, the proof-of-concept experiments are conducted, whose results demonstrate that it's effective to enhance the signal-to-noise ratio of rotational Doppler frequency spectrum and increase the distance of telemetry by employing EOV as the probe beam in the non-coaxial case. Meanwhile, the correctness and feasibility of the tilt angle measurement scheme of the object's rotating axis based on the EOVs with continuously varying ellipticities are proved. Compared to circular vortex beams, our scheme is able to improve signal-to-noise ratio and successfully measure the tilt angle of an object, which will help to expand the application scenarios of RDE and facilitate the engineering of rotational speed measurement.

## 2. Theory

### 2.1 Theoretical analysis and design of superposed elliptical vortex beams

Vortex beams are a class of structured beams with the helical phase $\exp(i\ell\varphi)$, whose expression can be summarized as $U(r,\varphi,z) = U_0(r,\varphi,z)\exp(-ikz)\exp(i\ell\varphi)$. As a type of OV, the vortex beam with Laguerre-Gauss (LG) mode in the cylindrical coordinate can be written as[25],

$$U_{OV}(r,\varphi,z) = U_z \cdot L_n^{|\ell|}(\frac{2r^2}{w_z^2})\exp(\frac{-ikr^2}{2R_z})\exp(i\ell\varphi), \tag{1}$$

where $U_z$ represents the electrical intensity at the propagation distance of $z$, $L_n^{|\ell|}(2r^2/w_z^2)$ denotes the Laguerre function, $k$ is the wave number, $\ell$ is the topological charge(TC), and $w_z$ indicates the waist radius at the propagation distance of z, which can be expressed as $2(z^2 + z_r^2)/kz_r$. $R_z$ is the radius of curvature of the wavefront and is written as $(z^2 + z_r^2)/z$. $z_r$ represents the Rayleigh radius. Firstly, we transform the polar coordinates $(r,\varphi)$ into Cartesian coordinates $(x_0, y_0)$. After that, by the coordinate stretching operation, the circular vortex beam can be converted into the EOV based on Eq.(2).

$$x = x_0,$$
$$my = y_0, \qquad (2)$$

where $m$ denotes the ellipticity, defined as the short half-axis length($b$) divided by the long half-axis length($a$) of the ellipse, i.e., $m = b/a, (1 \geq m \geq 0)$. The EOV can be expressed as,

$$U_{EOV}(x,y,z) = U_z L_n^{|\ell|}(\frac{2x^2 + 2m^2y^2}{w_z^2})\exp(\frac{-ik(x^2 + m^2y^2)}{2R_z})\exp(i\ell \tan^{-1}(\frac{my}{x})). \quad (3)$$

Based on the elliptical polar coordinates

$$\begin{cases} r_1 = \sqrt{\frac{x^2}{m^2} + y^2} \\ \psi = \tan^{-1}(\frac{my}{x}) \end{cases}, \qquad (4)$$

Eq.(3) can be written as,

$$U_{Eov}^{LG}(r_1,\psi,z) = U_z L_n^{|\ell|}(\frac{2m^2 r_1^2}{w_z^2})\exp(\frac{-ik(m^2 r_1^2)}{2R_z})\exp(i\ell\psi). \quad (5)$$

We simulate the intensity and phase profiles of the LG-mode EOVs of single and superimposed states, as shown in Fig. 1. According to Eq.(5), The spatially varying phase of the EOV is $\Phi = \ell\psi$, where $\psi$ is the azimuth angle in the elliptical polar coordinates. While in the circular coordinates, the azimuth angle $\theta$ can be expressed as $\theta = \tan^{-1}(y/x)$. Therefore, the relationship between azimuth angles in the two coordinate systems is,

$$\tan\psi = m\tan\theta \qquad (6)$$

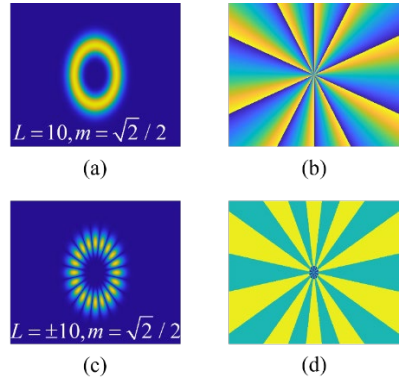

Fig. 1. Intensity and phase profiles of the LG-mode EOVs of single and superimposed states. (a), (b) The intensity and phase profiles of the EOV with TC($\ell$) of 10 and ellipticity($m$) of $\sqrt{2}/2$. (c), (d) The intensity and phase profiles of the EOV with TC($\ell$) of $\pm 10$ and ellipticity($m$) of $\sqrt{2}/2$.

## 2.2 The RDE mechanism of the elliptical optical vortex under the condition of tilted incidence

Consider a probe beam that can be written as $E(x,y,t) = E_0 \exp(-iw_0 t + ikz(t)) \exp(-i\Phi(x(t),y(t)))$. $\Phi(x(t),y(t))$ indicates the spatial phase. After the target modulation and the beat frequency processing, we can obtain the frequency shift of the optical echo signal, which can be expressed as,

$$f = \frac{2f_0 v_z}{c} + \frac{1}{2\pi}\nabla\Phi \bullet \vec{v}_{(x,y)}, \qquad (7)$$

where $c$ and $f_0$ are the speed and the frequency of light respectively. $\nabla\Phi$ represents the phase transversal gradient, $\vec{v}_{(x,y)}$ is the velocity of the scattered particle in the transverse plane of the object. The first term in Eq.(7) represents the linear Doppler frequency shift, and the second term is the frequency shift caused by the motion of the scattered particle in the transverse plane. When the probe beam is the OV with the spiral phase, the second term is converted into the RDE frequency shift, i.e., $f_{RDE} = \nabla\Phi \bullet \vec{v}_{(x,y)} / 2\pi$.

When the LG-mode EOV is employed to illuminate the oblique object with an angular velocity $\Omega$, the corresponding spatial relationship between the object and the probe beam is established, as shown in Fig. 2. Fig. 2 (a) is the detection schematic, and (b) is the geometric model of the RDE of the EOV. We establish the spatial rectangular coordinate system with the center($O$) of the EOV as the origin. The propagation direction of beam is the z-axis, and the horizontal direction in the transverse plane of the beam is the y-axis. $A$ is the scattered particle on the EOV field, $\vec{r}$ represents the radial vector from the origin to the scattered particle($A$), and $\vec{\omega}$ is the angular velocity vector of the object, whose direction is normal to the plane of the rotating object. The magnitude of $\vec{\omega}$ is $\Omega$. We decompose the vector $\vec{\omega}$ orthogonally along the coordinate axis. The angle between the vector $\vec{\omega}$ and the z-axis is $\beta$ which denotes the tilted angle, and the angle between the projection of $\vec{\omega}$ on the $xoy$ plane and the x-axis is $\gamma$. $\vec{V}_A$ is the linear velocity at the scattering point $A$.

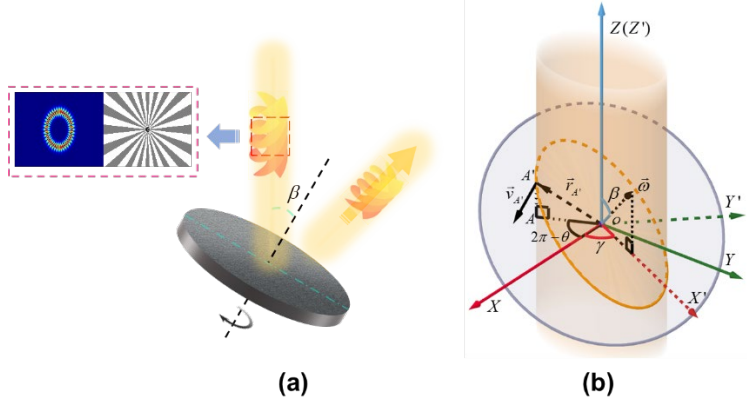

Fig. 2. Schematic of detection based on the EOV and geometric model of non-coaxial RDE. (a) Schematic of the EOV obliquely illuminating a rotating object. $\beta$ indicates the tilt angle .(b) Geometric model of the elliptical vortex beam for RDE derived from (a).

In order to clearly reveal the tilted RDE mechanism of the EOV and illustrate the effectiveness of the EOV-based RDE signal correction method at oblique incidence, we further analyze the rotational Doppler shift of each scattering point based on geometric model. In fact, the total rotational Doppler frequency shift of the echo signal is the superposition of the frequency shifts of each scattering points in the optical field. Based on the vector relationship in the coordinate system, the relationship among $\vec{\omega}$, $\vec{r}$ and $\vec{V}_A$ can be derived,

$$\begin{aligned} \vec{V}_A &= \vec{\omega} \times \vec{r} \\ \vec{\omega} \cdot \vec{r} &= 0 \end{aligned} \tag{8}$$

based on the relationship between $\psi$ and $\theta$ in Eq.(6),

$$\vec{r} = mr\left(\frac{1}{\sqrt{1+m^2\tan^2\theta}}, \frac{\tan\theta}{\sqrt{1+m^2\tan^2\theta}}, \frac{-\tan\beta\cos\gamma - \tan\beta\sin\gamma\tan\theta}{\sqrt{1+m^2\tan^2\theta}}\right). \tag{9}$$

Therefore, the expression of the linear velocity $\vec{V}_A$ at point $A$ can be deduced as,

$$\begin{aligned} \vec{V}_A &= \vec{\omega} \times \vec{r} \\ &= \frac{\Omega m r}{\sqrt{1+m^2\tan^2\theta}} \cdot (P\vec{e}_x + Q\vec{e}_y + J\vec{e}_z) \end{aligned} \tag{10}$$

where

$$\begin{aligned} P &= -\tan\beta\cos\gamma\sin\beta\sin\gamma - \tan\beta\sin^2\gamma\tan\theta\sin\beta - \cos\beta\tan\theta \\ Q &= \cos\beta + \sin\beta\cos^2\gamma\tan\beta + \sin\beta\cos\gamma\tan\beta\sin\gamma\tan\theta \\ J &= \sin\beta\cos\gamma\tan\theta - \sin\beta\sin\gamma \end{aligned} \tag{11}$$

Thus, the linear velocity at the scattering point $A$ can be expressed in the spatial rectangular coordinate as $\vec{V}_A = \Omega m r(P,Q,J)/\sqrt{1+m^2\tan^2\theta}$. It is shown that the structural phase of EOV is $\Phi = \ell\psi$ in Eq.(5). Combining with Eq.(6), the phase transversal gradient of EOV can be written as,

$$\nabla\Phi = \frac{\ell}{r}\left(\frac{-\tan\theta}{\sqrt{1+m^2\tan^2\theta}}, \frac{1}{\sqrt{1+m^2\tan^2\theta}}, 0\right). \tag{12}$$

Then, the total rotational Doppler shift generated by the EOV under tilted incidence condition can be given by,

$$\begin{aligned} f_{EOV} &= \frac{1}{2\pi}\nabla\Phi \cdot \vec{V}_A \\ &= \frac{m\ell\Omega}{2\pi(\cos^2\theta + m^2\sin^2\theta)} \cdot [\cos\beta + \tan\beta\sin\beta\cos^2(\theta-\gamma)] \end{aligned} \tag{13}$$

From Eq.(14), we can see that the rotational Doppler shift generated by the EOV is not only related to the rotating speed ($\Omega$) of the object and the TC ($\ell$) of the EOV, but also to the ellipticity ($m$) of the EOV and the spatial azimuth ($\beta,\gamma$) of the object. This provides a powerful way for us to use the parameters of the EOV (such as ellipticity) to correct the rotational Doppler broadening effect and measure the tilt angle. When the circular-OV is employed as the probe light[26], the RDE spectrum from the echo will be broadened with the frequency shift range of $[\ell\Omega/(2\pi\cdot\cos\beta), \ell\Omega\cos\beta/2\pi]$ in the tilted case. $\beta$ represents the tilt angle. However, when the long axis of the EOV is in the same direction as the inclination of the object, i.e., $\gamma = 0$ and the ellipticity of the EOV corresponds strictly to the tilt angle of the object which means $m = \cos\beta$ is satisfied, The rotational Doppler shift formula (14) of EOV degenerates to $f_{EOV} = \ell\Omega/2\pi$, that is, the RDE frequency spectrum does not expand. It can be seen from the above analysis that when an EOV with proper ellipticity is employed as the probe beam, the defects of signal-to-noise ratio reduction and spectral broadening caused by tilted incidence can be overcome entirely in theory, so that the rotational Doppler shift become an individual peak. This contributes to improve the detection distance and the measurement accuracy.

Based on Eq. (14), we simulate the rotational Doppler shift distribution in the case of different tilt angle($\beta$) of the object and different ellipticity($m$) of the EOVs. Here, some parameters need to be defined firstly. We set the TC of EOV to $\pm 20$, the rotating speed of the object to 50Hz, $\gamma = 0$, and the tilt angle range of the object to $[0, 2\pi/5]$. Then the rotational Doppler shift of the scattered light is simulated numerically. The results are shown in Fig. 3.

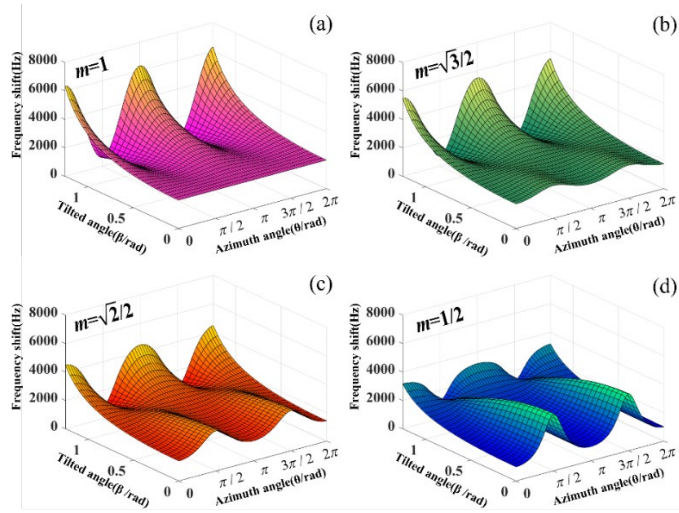

Fig. 3. RDE simulation results of the EOVs in the case of different tilt angles. (a)-(d) Rotational Doppler shift varying with tilt angle($\beta$) based on the EOV with TC of $\pm 20$ and ellipticity($m$) of $1, \sqrt{3}/2, \sqrt{2}/2, 1/2$ in turn. The rotating speed of the object in the simulation is set to 50 Hz. When $m = \cos\beta$, the rotational Doppler shift is a constant, 2000Hz.

The x-axis indicates the azimuthal angle($\theta$), the y-axis represents the tilt angle($\beta$) of the object, and the z-axis is the rotational Doppler shift generated by the EOV in Fig. 3. Fig. 3 (a)-(d) represent the simulation results of rotational Doppler shift that varies with the tilt angle of the object and azimuth angle when the ellipticity of EOV is $1, \sqrt{3}/2, \sqrt{2}/2, 1/2$ respectively. It can be seen from Fig. 3 that when the tilt angle of the object is $\pi/6$, $\pi/4$, $\pi/3$ in turn and $m = \cos\beta$ is met, the rotational Doppler shift caused by EOVs with different ellipticities is constant and the frequency spectrum is no longer broadened. This is in good agreement with the theoretical analysis. Moreover, according to the Fig. 3, even if the ellipticity of EOV does not correspond exactly to the tilt angle of the object, the corrective method based on the EOV is still effective and can partially eliminate the frequency spectral expansion introduced by oblique incidence within a certain fluctuating range.

In practical measurement, since the tilt angle of the object is unknown, we can use a series of EOVs with continuously varying ellipticities to illuminate the rotating object and collect the echo signal. Then the time-frequency transform method is employed to process the time-domain optical signal. By analyzing the time-frequency spectrum, the ellipticity of the EOV corresponding to the narrowest spectral expansion can be obtained. When the ellipticity of the EOV corresponds strictly to the tilt angle of the object( $m = \cos\beta$ ), the rotational Doppler spectrum hardly broadens and the signal-to-noise ratio is maximum. Then, we are able to get the inclined angle of the object. The above theoretical analysis and simulation results show that compared with the conventional circular OV as the probe beam, the correction method based on the EOV may help to improve the signal-to-noise ratio and enhance the measurement accuracy of rotating speed and detection distance in the case of tilt incidence. Also, by employing a set of EOVs with different ellipticities, the oblique angle of the object can be measured.

## 3. Experiments and discussion

A proof-of-concept experiment is designed to demonstrate the correctness of the above theoretical analysis. The experimental setup is shown in Fig. 4. Since the SEOV has self-interference characteristics that can omit the reference optical branch and reduce the experimental requirements, we use the complex amplitude modulation method to generate the holograms of the SEOVs with different ellipticities based on Eq.(3). The linearly polarized laser beam with a wavelength of 532nm illuminates the spatial light modulator (SLM) with the phase holograms of the SEOVs. The polarizer(Pol) controls the beam to be horizontally polarized. The first-order diffracted light is selected by L3, Aperture (Ap), and L4. Afterwards, the beam is divided into two paths by the beam splitter(BS), one path is captured by the CCD to analyze the beam profile, and the other path is used to irradiate the tilted rotor. Finally, the diffused light is captured by the lens(L5) and converged to the avalanche photodetector(APD). The rotor is about 815mm away from the SLM. The computer in Fig. 4 is employed to process the optical signal collected by the APD and modulate the holograms.

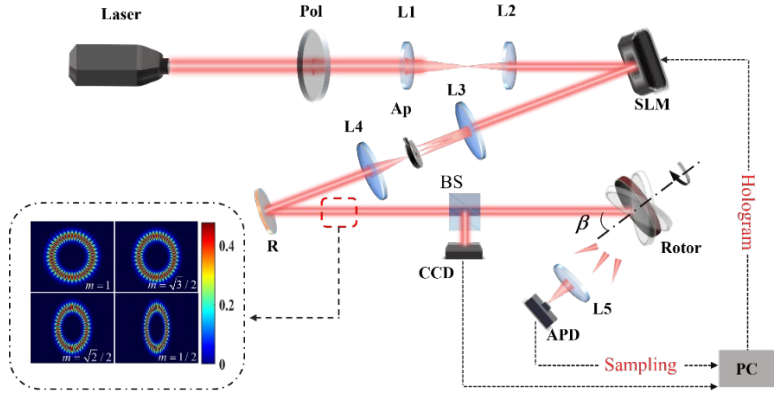

Fig. 4. Experimental setup for oblique irradiation of the EOV. Pol, polarizer. L1, L2, L5,lens with a focal length of 75mm. L3, L4, lens with a focal length of 50mm. SLM, spatial light modulator. Ap, Aperture. R, reflector. BS, beam splitter. APD, avalanche photodetector. $\beta$ indicates the tilt angle .The PC is used to process the data and modify the holograms. The intensity profiles of the employed probe beams are shown on the left of the graph.

Firstly, we set the TC of LG-mode superposed EOV to $\pm 20$, the rotating speed of the object to 50Hz, and the angle $\gamma$ to 0 in the experiment. Then, we employed the SEOVs with ellipticities of $1$, $\sqrt{3}/2$, $\sqrt{2}/2$, and $1/2$ to illuminate the rotating object with different tilt angles. And when the ellipticity is 1, The SEOV becomes the conventional circular superposition-mode OV whose measurement results are treated as the reference group. The tilt angles( $\beta$ ) were set to 0, $\pi/6$, $\pi/4$, and $\pi/3$ in turn during the experiment. Finally, we used APD to capture the time-domain echo signal within 1s. After Fourier transform, the final experimental results are shown in Fig. 5.

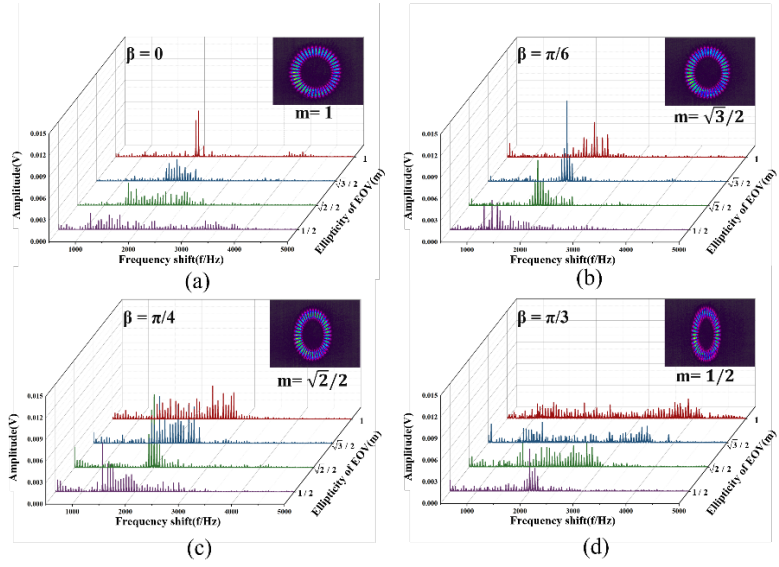

Fig. 5. Experimental results of EOV-based RDE at different tilt angles($\beta$) of the object.(a)-(d) The rotational Doppler shift generated by the EOVs with different ellipticities when the tilt angles of the object are 0, $\pi/6$, $\pi/4$, and $\pi/3$ in order. The rotational Doppler signal in frequency-domain has the highest amplitude and almost no expansion when $m = \cos\beta$.

Fig. 5 represents the rotational Doppler shift obtained by irradiating the rotating object with the SEOVs of different ellipticities when the tilt angle of the rotor is 0, $\pi/6$, $\pi/4$, and $\pi/3$, respectively. At the object's tilt angle of $\pi/6$, the rotational Doppler spectrum is obviously broadened and the frequency-domain peak of the Doppler echo signal is 0.004V if the SEOV with $m = 1$ is employed as the probe beam, as shown by the red line in Fig. 5 (b). However, when the probe beam is the SEOV with the ellipticity of $\sqrt{3}/2$, the Doppler spectrum hardly broadens, and the amplitude of the Doppler echo signal in frequency-domain is 0.011V which is about three times of that before the correction, as shown by the blue line in Fig. 5 (b). The peaks of the rotational Doppler frequency spectrum generated by the superposed EOVs with appropriate ellipticities($m = \cos\beta$) are increased by 0.006V and 0.003V compared to that generated by the circular OV when the tilt angle is $\pi/4$ and $\pi/3$, respectively. And the amplitudes of the rotational Doppler spectrum increase about twice as much as before the correction when the tilt angles are $\pi/4$ and $\pi/3$. The experimental results show that by the correction method of rotational Doppler shift at tilt incidence in this paper, the spectral expanding effect caused by the inclination is eliminated basically and the amplitude of the rotational Doppler shift is greatly enhanced. Note that although the correction effect is the best when the tilt angle and the ellipticity of EOV satisfy the relation $m = \cos\beta$, the method can still optimize the RDE signal to a certain extent when the tilt angle and the ellipticity are within a fixed range of deviation.

To further prove the correctness of the theoretical model of the RDE based on the EOV, we analyze the error between the above experimental data and theory. the results are shown in Fig. 6.

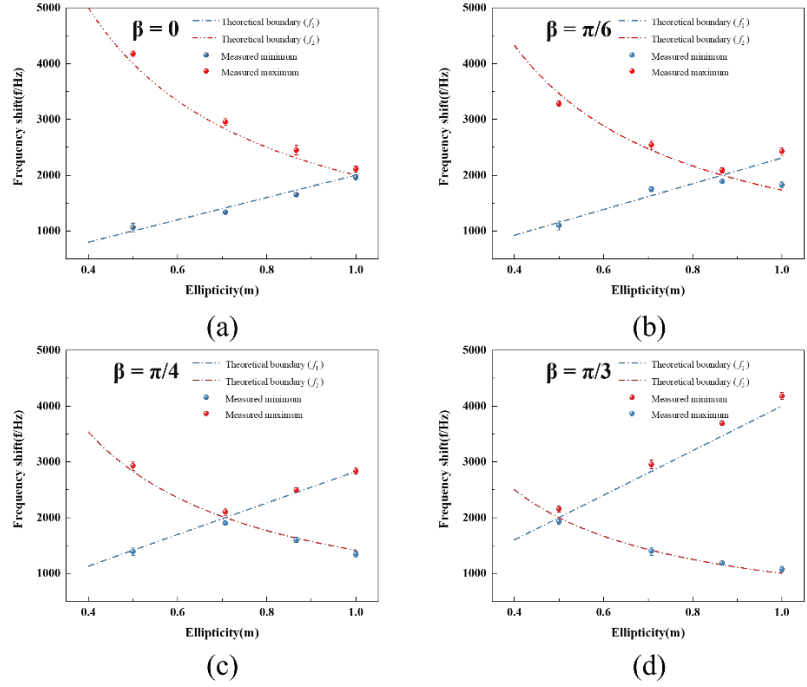

Fig. 6. Experimental and theoretical results of the boundary values of the rotational Doppler spectral range.(a)-(d) The broadening range of rotational Doppler shift varies with ellipticity of the EOV when the tilt angle is 0, $\pi/6$, $\pi/4$, and $\pi/3$ successively. The red and blue lines represent the limits of the rotational Doppler spectrum in theory. The area between the two lines represents the spreading range of the rotational Doppler spectrum. The red and blue dots indicate the maximum and minimum values of the rotational Doppler shift in the experiment, respectively. The experimental results are in good agreement with the theory.

Fig. 6 (a)-(d) indicate the boundary of the broadened rotational Doppler frequency spectrums generated by the SEOVs with different ellipticities at different tilt angles of the object, respectively. In order to detect the boundary of the rotational Doppler spectrum, we regard the signal region greater than 7 times the average signal amplitude within [1000Hz,5000Hz] in the frequency domain as the rotational Doppler spectrum range. The dashed lines represent the theoretical maximum and minimum values of the rotational Doppler frequency spectrum range generated by the SEOVs based on Eq.(14). The area between the two dashed lines represents the rotational Doppler spectral spreading range. Taking Fig. 6 (a) as an example, we measured the maximum and minimum rotational Doppler shifts corresponding to the superposed EOVs with the ellipticity of $1, \sqrt{3}/2, \sqrt{2}/2$, and $1/2$ when the tilt angle of the object is 0, respectively. And the rotational Doppler spectrum is the narrowest, and the amplitude is the maximum when the ellipticity of the probe beam is 1. The same is true for (b)-(d) in Fig. 6. The analysis results indicate that although the experimental results are in good agreement with the theory, there are still some errors. The maximum measurement errors in Fig. 6 (a)-(d) are 6.2%, 6.9%, 5.2%, and 7.6% in Fig. 6 (a)-(d) when the tilt angles($\beta$) are 0, $\pi/6$, $\pi/4$, and $\pi/3$, respectively. These errors are caused by various factors, such as the vibration of the rotor, the relative spatial relationship between the beam and the object, and the thermal noise, etc. Of these, the mode purity of SEOV has a significant impact on the experimental results and is the main factor to introduce extra sidebands for rotational Doppler shift. This is also why the rotational Doppler shift is not a single peak even if the ellipticity of the EOV satisfies the relationship $m = \cos\beta$ with the tilt angle of the object. In summary, the experimental results agree well with the theoretical RDE mechanism of EOV within a certain error range, which demonstrates the effectiveness of the corrective method about the rotational Doppler shift based on the EOV under the tilt condition. The scheme improves the signal-to-noise ratio and broadens the scope of RDE applications. In addition, since tilt illumination is the more general case in practical measurement, this method has great potential for application in telemetry.

Furthermore, we cannot determine the optimal ellipticity of the SEOV for correction because the inclined angle of an object is unknown in actual measurement. Therefore, we employ a set of SEOVs with continuously varying ellipticities to illuminate a tilted rotating object and receive the optical echo including at least one period. The measurement scheme is shown in the Fig. 7. Then, through the time-frequency analysis method, the measurement results displayed in the time-frequency domain can be obtained. Since the ellipticity of the probe beam is related to the switching time, and the time is connected with the rotational Doppler shift in the time-frequency spectrum, we can obtain the correlation between the ellipticity and the rotational Doppler shift. When the ellipticity of the EOV and the tilt angle of the object strictly meet the relationship $m = \cos\beta$, the rotational Doppler spectrum hardly broadens, and the signal amplitude is the maximum in the time-frequency domain. The relationship between the tilt angle ($\beta$) and the corresponding time ($T_{min}$) at this moment can be expressed by,

$$\beta = (\frac{T_{na} - T_{in}}{\Delta T} + 1) \cdot \cos^{-1}(\Delta m), \tag{14}$$

where $T_{na}$ is the corresponding time when the rotational Doppler spectrum is the narrowest, $T_{in}$ is the initial time of one period, $\Delta T$ is switching interval of the holograms, and $\Delta m$ is the ellipticity interval of SEOVs. By this method, we can simultaneously determine the tilt angle of the object and the optimal ellipticity of the SEOV.

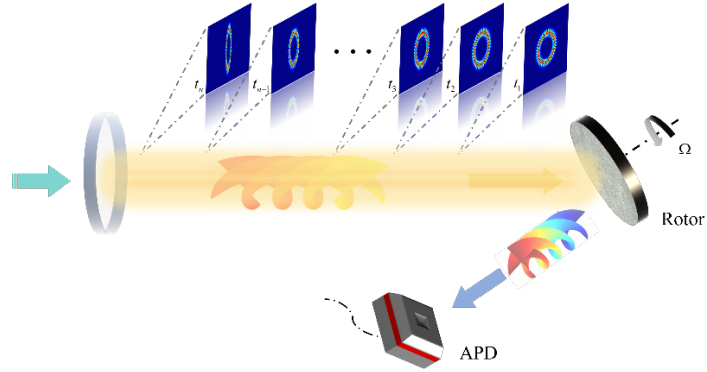

Fig. 7. Schematic of the measurement scheme of the tilt angle and the ellipticity of the corrected EOV for an unknown object. A set of EOVs with continuously varying ellipticity continuously illuminates the rotating object. The detector receives optical echo from the object within at least one period. $t_i$ represents the switching time of the EOV with different ellipticity.

Another experiment is conducted to prove the correctness of the above method, the experimental setup remains unchanged as shown in the Fig. 4. Firstly, we use the complex amplitude modulation method to modify the ellipticity of the SEOV, and design a group of SEOV holograms with the ellipticity range of $[\cos(11\pi/24), 1]$ and an interval ($\Delta m$) of $\cos(\pi/72)$. This group contains 34 holograms of SEOV. They are sequentially displayed on the SLM to obtain a series of SEOVs with successively varying ellipticities. We set the rotating speed($\Omega$) to 50Hz, the tilt angle($\beta$) to $\pi/4$, and the TC to $\pm 20$. The total cycle period($T$) of a group of SEOVs is set to 1.7s ($T=1.7s$), which means the display time ($\Delta T$) of each hologram is 0.05s. The rest of the experimental conditions are the same as the previous experiment. After the light-matter interaction, we collect the optical echo from the object within 2 s (containing only one cycle), whose time-domain distribution is shown in Fig. 8 (a). Afterwards, the short-time Fourier transform(STFT) is employed to obtain the time-frequency spectrum as shown in Fig. 8 (b). It can be seen that the rotational Doppler frequency band firstly shrinks and then broadens with time or the ellipticity of the beam, and the intensity in the frequency domain is strongest at the narrowest part of the band. To further measure the optimal ellipticity and tilt angle, a filtering algorithm is designed to remove noise from the time-frequency domain signal, and the result is shown in Fig. 8 (c). As the time-frequency spectrum is affected by the window length, sampling time, sampling frequency and other factors, it is difficult to get the precise tilt angle of the object and the ellipticity of the beam as accurately as the theoretical analysis. Therefore, we extract the upper edge of the time-frequency spectrum and use the trigonometric functions to fit the edge to obtain the minimum on the curve, which corresponds to the narrowest RDE spectrum, as shown in Fig. 8 (d). Finally, the measurement results of the initial time($T_{in}$) and the optimal time($T_{na}$) are 0.027s, 0.818s, respectively. the ellipticity of SEOV is 0.74 approximately and the tilt angle of the object is 42.05 degree based on Eq.(15) and $m=\cos\beta$. The measurement error is 6.5%. The maximum tilt angle we can measure under the above experimental conditions is 75 degree. This experimental result demonstrates that the simultaneous measurement of the tilt angle of the object and the optimal ellipticity of the EOV for the correction can be achieved by this method.

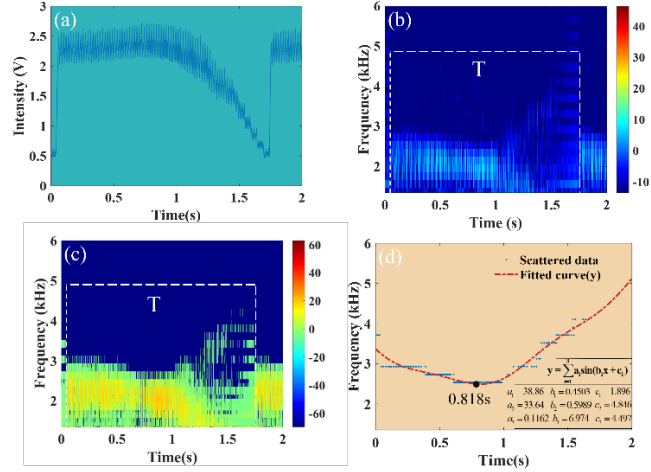

Fig. 8. Experimental results of a set of EOVs with different ellipticities irradiating a rotating object with unknown tilt angle. (a) the original data in time domain within 2s. (b) The time-frequency domain spectrum of the signal after STFT. T is the detection period of a set of SEOVs. T=1.7s. (c) Time-frequency spectrum after filtering. (d) The curve fitting results after the upper edge of the time-frequency spectrum is extracted. 0.818s is the horizontal coordinate of the lowest value of the fitted curve, which represents the time when the Doppler spectrum spreading range is minimum. The range and interval of ellipticity are $[\cos(11\pi/24),1]$, $\cos(\pi/72)$. The switching interval of EOV is 0.05s.

Note that ,the resolution of the measured obliquity of the object is $\pi/72$ since the ellipticity interval of the probe beam is $\cos(\pi/72)$. If more accurate measurement is desired, the ellipticity interval between SEOVs should be shortened. But it also means that the longer sampling time is required. In addition, the rotational Doppler shift is not a single peak as the theory when the equation $m = \cos\beta$ is met in the experiment due to the mode purity, vibration of the rotor and the interference of the environmental noise, etc.

Also, it is known that the LG-mode EOV turns out to be rotated by 90 deg with respect to the initial beam at $z = 0$ when the propagation distance is far enough according to the propagation characteristics of the beam. Therefore, we can correct the rotating angle of the EOV in advance for telemetry. In remote sensing, we can directly set the corrected rotating Angle to $\pi/2$.

## 4. Conclusion

In summary, we reveal the RDE mechanism of the EOV under three-dimensional conditions and propose a rotational Doppler correction method based on EOV in the more general case. In this scheme, by employing the EOV with proper ellipticity to detect a tilted rotating object, we can reduce the broadening range of rotational Doppler frequency spectrum and enhance the signal-to-noise ratio of the optical echo. Compared with the conventional circular OV, when the EOV whose ellipticity met the equation $m = \cos\beta$ with inclined angle is used as probe beam, the rotational Doppler spectral broadening effect can be completely removed in theory. This method is helpful to broaden the scope of RDE application and increase the detection distance, which may have important application value in telemetry.

In addition, considering the case where the tilt angle of the object is unknown, we use a set of SEOVs with continuously varying ellipticity to irradiate the tilted rotating object. By analyzing the time-frequency spectrum of the echo signal, we can obtain the time corresponding to the minimum of the rotational Doppler frequency spectral spreading range. At this moment, the tilt angle is strictly correlated with the ellipticity. Since the switching time is related to the beam ellipticity, the tilt angle of the object and the optimal ellipticity can be obtained simultaneously by our scheme, which contributes to the acquisition of more information about the orientation of an object and the engineering of rotational measurement. Also, we consider the propagation of the EOV. According to the propagation characteristics of SEOV, we can correct the rotational angle of SEOV in advance in the actual measurement.